\documentclass[10pt]{amsart}
\usepackage{epsfig,url}
\usepackage{mathdots}

\theoremstyle{definition}

\theoremstyle{remark}

\begin{document}

\title[Analytic expressions for Debye functions]
{Analytic expressions for Debye functions and the heat capacity of a solid}

\author[Ivan Gonzalez et al]
{Ivan Gonzalez}
\address{Departmento de F\'{i}sica y Astronomia,
Universidad de Valparaiso, Valparaiso, Chile}
\email{ivan.gonzalez@uv.cl}

\author[]
{Igor Kondrashuk}
\address{Grupo de Matem\'{a}tica {A}plicada {\rm \&} Grupo de F\'isica de Altas Energ\'ias, \\
Departmento de Ciencias B\'{a}sicas,
Universidad del B\'{i}o-B\'{i}o, Campus 
Fernando May, Casilla 447, Chill\'{a}n, Chile}
\email{igor.kondrashuk@gmail.com}

\author[]{Victor H. Moll}
\address{Department of Mathematics,
Tulane University, New Orleans, LA 70118}
\email{vhm@tulane.edu}

\author[]{Alfredo Vega}
\address{Departmento de F\'{i}sica y Astronomia,
Universidad de Valparaiso, Valparaiso, Chile}
\email{alfredo.vega@uv.cl}


\date{\today}


\begin{abstract}
Analytic expressions for the $N$-dimensional Debye function are obtained by the method of brackets. The new 
expressions are suitable for the analysis of the asymptotic behavior of this function, both in the high and low 
temperature limits. 
\end{abstract}

\maketitle

\newcommand{\ba}{\begin{eqnarray}}
\newcommand{\ea}{\end{eqnarray}}
\newcommand{\ift}{\int_{0}^{\infty}}
\newcommand{\nn}{\nonumber}
\newcommand{\no}{\noindent}
\newcommand{\lf}{\left\lfloor}
\newcommand{\rf}{\right\rfloor}
\newcommand{\realpart}{\mathop{\rm Re}\nolimits}
\newcommand{\imagpart}{\mathop{\rm Im}\nolimits}

\newcommand{\op}[1]{\ensuremath{\operatorname{#1}}}
\newcommand{\pFq}[5]{\ensuremath{{}_{#1}F_{#2} \left( \genfrac{}{}{0pt}{}{#3}
{#4} \bigg| {#5} \right)}}

\newtheorem{Definition}{\bf Definition}[section]
\newtheorem{Thm}[Definition]{\bf Theorem}
\newtheorem{Example}[Definition]{\bf Example}
\newtheorem{Lem}[Definition]{\bf Lemma}
\newtheorem{Cor}[Definition]{\bf Corollary}
\newtheorem{Prop}[Definition]{\bf Proposition}
\numberwithin{equation}{section}

\section{Introduction}
\label{sec-intro}

The $N$ dimensional Debye functions play an important role in study of a variety of 
problems in statistical physics and solid state physics, especially in
calculations of heat capacity of solids. This function appeared first in
a model proposed by Debye \cite{debye-1912a} describing  the heat capacity of a
crystalline solid, which with some variations it is still used today.

These functions, up to recent times only known through their
integral representation, have created enough interest in their evaluation for arbitrary 
values of  $N$ and absolute
temperature $T$. Debye functions can be expressed as \cite[ch.~27]{abramowitz-1972a}:

\begin{eqnarray}
D_{N}( X) &   = & \frac{N}{X^{N}} \int_{0}^{X}\frac{t^{N}}{e^{t}-1}dt  \label{debye-1} \\
&  = &  N \left( \frac{1}{N}-\frac{X}{2\left( N+1\right) }+\sum_{k=0}^{\infty}
\frac{B_{2k}}{\left( 2k+N\right) \Gamma \left( 2k+1\right) }
X^{2k}\right), \nonumber
\end{eqnarray}
\noindent
for $|X| <2\pi$ and $N \geq 1$.  Here $B_{k}$ are the  Bernoulli numbers. Numerical computation of 
these functions appear in \cite{huang-1975a,mcquarrie-1975a}. The  first
analytical expression for such functions, other than integral representations, may be found  in 
\cite{dubinov-2008a}. The work presented here gives expressions for $D_{N}(X)$
in terms of the polylogarithm functions.

The  \texttt{method of brackets} \cite{gonzalez-2010a} is used in the current work  to evaluate
the Debye functions. This method gives (new) analytic expressions for 
them, reproducing the results presented in \cite{dubinov-2008a}, as well as some  new expressions.

The method of brackets was developed in the context of  calculations of  multidimensional definite integrals
appearing  in evaluation of Feynman diagrams \cite{gonzalez-2010d,gonzalez-2007a, gonzalez-2008a, gonzalez-2009a}. It 
consists in a small number of heuristic rules that  yield the evaluation 
of a  wide range of integrals. These rules admit an easy 
implementation in a computer algebra system. The reader will find more details  in \cite
{amdeberhan-2012b,gonzalez-2014a,gonzalez-2010a,gonzalez-2015b,gonzalez-2010b}.

The content of the paper is described next. Section \ref{sec-2} introduces the method of brackets, 
Section \ref{sec-3} uses this method to evaluate Debye functions and their analytic expressions. In 
particular, expressions that are free of integral representations are presented here, recovering those presented by \cite{dubinov-2008a}. These results are then used to 
study the asymptotic behaviour of these functions in limiting values of the  temperature. Section \ref{sec-4} uses these 
representations to evaluate the internal
energy and heat capacity in solids. The emphasis here is on the new expression
for Debye functions  to show that the 
manipulation of them simplifies the computation of the  limits $T\rightarrow 0$ and $T\rightarrow
\infty$ for temperature presented in \cite{dubinov-2008a}.

\section{Basic of Method of Brackets (MoB)}
\label{sec-2}

The method of brackets is a generalized version of the \texttt{Negative
Dimensional Integration Method }(NDIM) \cite{anastasiou-2000a, anastasiou-2000b,
dunne-1987a, halliday-1987a, suzuki-2002a}, a technique developed to evaluate Feynman diagrams. In quantum
field theories, Feynman diagrams correspond to multi-variable integrals that
represent physical processes. 

This method evaluates definite  integrals in one or several dimension over the 
interval  $\left[ 0,\infty \right]$. The procedure introduces the notion of a bracket and 
converts the integrand in a series of brackets. The method contains a small number of 
heuristic rules which transform the evaluation of an integral into the solution of a 
small linear system of equations. A summary of these rules is presented below. More 
details may be found in \cite{amdeberhan-2012b,gonzalez-2014a,gonzalez-2010a,gonzalez-2010b}.

\smallskip 

\noindent
\texttt{Rule 0}. For $a \in \mathbb{C}$, the \textit{bracket} associated to 
$a$ is the divergent integral
\begin{equation}
\langle a \rangle = \int_{0}^{\infty} x^{a-1} \, dx.
\end{equation}

\smallskip

\noindent
\texttt{Rule 1}. \textit{The expansion of an  arbitrary function}. The use of the method of 
brackets requires to replace components of the integrand by their 
his corresponding power series, that is, it is required to represent an
arbitrary function $f(x)$ as:
\begin{equation}
f(x) =\sum\limits_{n}\phi_{n}C(n)  x^{\beta n+\alpha },
\end{equation}
\noindent
where $C(n)$ are the coefficients in the expansion, $\alpha$ and $\beta $ are arbitrary (complex) exponents and $\phi _{n}$ is
defined by:
\begin{equation}
\phi _{n}=\frac{(-1)^{n}}{\Gamma(n+1)}.
\end{equation}

For multidimensional integrals one needs expansions in several variables, such as 
\begin{equation}
f( x_{1},x_{2}) =\sum\limits_{n_{1}}\sum\limits_{n_{2}}\phi
_{n_{1}}\phi _{n_{2}}\;C( n_{1},n_{2}) \;x_{1}^{\beta_{1}n_{1}+\alpha _{1}}x_{2}^{\beta _{2}n_{2}+\alpha _{2}}.
\end{equation}
\noindent
The notation $\phi_{12}$ is frequently used for $\phi_{n_{1}} \phi_{n_{2}}$. 

\smallskip

\noindent
\texttt{Rule 2}. \textit{Polynomial expansion}.  An expression of the form $( A_{1}+ \cdots +A_{r})^{\mu}$ often 
appears in the evaluation of integrals. The expansion
\begin{equation}
(A_{1}+\cdots+A_{r}) ^{\mu}=\sum \limits_{n_{1}}...\sum
\limits_{n_{r}}\phi_{n_{1}}...\phi_{n_{r}}\; A_{1}^{n_{1}} \cdots A_{r}^{n_{r}} \frac{\left\langle
-\mu+n_{1}+...+n_{r}\right\rangle }{\Gamma( - \mu) }.
\nonumber
\end{equation}
\noindent
This rule has been established in \cite{gonzalez-2010a}.

\smallskip

\noindent
\texttt{Rule} $3$: \textit{Eliminating integration symbols}. Once the first two rules are applied, the integral is converted into 
a bracket series using the definition of bracket. 

\smallskip

\noindent
\texttt{Rule} $4$: \textit{Finding solutions}.  The result of applying the previous rules to an integral is that its value is 
 represented by a bracket series $J$.  The rule to evaluate this series is given in the special case when number of sums
 and brackets is the same (this is the so-called \textit{index zero case}): the bracket series is
\begin{equation*}
J=\sum\limits_{n_{1}} \cdots \sum\limits_{n_{r}}\phi _{n_{1}}...\phi
_{n_{r}}\;C(n_{1},\cdots ,n_{r})\langle a_{11}n_{1}+ \cdots +a_{1r}n_{r}+c_{1}\rangle
\cdots \langle a_{r1}n_{1}+ \cdots +a_{rr}n_{r}+c_{r}\rangle. 
\end{equation*}
The coefficient $C(n_{1},...,n_{r})$ depends on the 
parameters of the integral and the index of the sum  $\left\{ n_{i}\right\}, \, i=1,...,r$.  The value of this multiple sum is 
declared to be
\begin{equation}
\mathbf{J}=\frac{1}{\left\vert \det \left( \mathbf{A}\right) \right\vert }
\;\Gamma \left( -n_{1}^{\ast }\right) \cdots \Gamma \left( -n_{r}^{\ast }\right)
\;C(n_{1}^{\ast },...,n_{r}^{\ast }),
\end{equation}
where $\mathbf{A} =  \{ a_{ij} \}$ and the values $\left\{ n_{i}^{\ast}\right\}$ $\left(i=1,...,r\right)$
are the  solutions of the linear system obtained by the vanishing of the brackets:
\begin{equation}
\left\{
\begin{array}{cc}
a_{11}n_{1}+...+a_{1r}n_{r}= & -c_{1} \\
\vdots & \vdots \\
a_{r1}n_{1}+...+a_{r}n_{r}= & -c_{r}.%
\end{array}%
\right.
\end{equation}

If the matrix $\mathbf{A}$ is not invertible and the  number of sums
is larger than the number of brackets, there is an extension of the procedure described here to evaluate the integral. 
Details may be found in \cite{gonzalez-2010a,gonzalez-2010b}.

\section{The Debye function $D_{N}(\alpha ,X)$}
\label{sec-3}

The Debye functions is defined by:
\begin{equation}
D_{N}(X) =\frac{N}{X^{N}}\int_{0}^{X}\frac{t^{N} \, dt}{e^{t}-1}.  \label{DebyeFunction}
\end{equation}
\noindent
The following extension is considered here:
\begin{equation}
D_{N}(\alpha ,X) =\frac{N}{X^{N}}\int_{0}^{X}\frac{t^{N} \, dt}{e^{t}-\alpha}.  \label{eqq3}
\end{equation}%
Here $N$ is zero or a positive integer, $X$ and $\alpha$ are positive parameters. The parameter $\alpha$
is introduced here to find alternative expressions for these extensions.

\subsection{A bracket series for $D_{N}(\alpha,X)$}

The computation of a bracket series for $D_{N}\left( \alpha ,X\right)$ is described next.  The first step is the 
expansion of the denominator in the integrand to obtain:
\begin{equation}
\label{dn-1}
D_{N}( \alpha ,X) =\frac{N}{X^{N}}\sum\limits_{n_{1}}\sum\limits_{n_{2}}\phi _{n_{1}}\phi _{n_{2}}\;\left( -1\right)^{n_{2}}\alpha
^{n_{2}}\langle 1+n_{1}+n_{2}\rangle \int_{0}^{X}t^{N}e^{tn_{1}} dt.
\end{equation}
\noindent
The expansion of the exponential function is 
\begin{equation}
e^{tn_{1}}=\sum\limits_{n_{3}}\frac{1}{n_{3}!}t^{n_{3}}n_{1}^{n_{3}}=\sum
\limits_{n_{3}}\phi _{n_{3}}\left( -1\right) ^{-n_{3}}t^{n_{3}}n_{1}^{n_{3}},
\end{equation}
and replacing  in \eqref{dn-1}  produces 
\begin{equation*}
D_{N}( \alpha ,X) =\frac{N}{X^{N}}\sum\limits_{n_{1}}\sum%
\limits_{n_{2}}\sum\limits_{n_{3}}\phi _{n_{1}}\phi _{n_{2}}\phi
_{n_{3}}\;\left( -1\right) ^{n_{2}-n_{3}}\alpha ^{n_{2}}n_{1}^{n_{3}}\langle
1+n_{1}+n_{2}\rangle \int_{0}^{X}t^{N+n_{3}}\;dt.
\end{equation*}
\noindent
The change of variables $y = t/(X-t)$ converts the last integral to $[0, \, \infty)$ as 
\begin{equation}
\int_{0}^{X}t^{N+n_{3}}\;dt=X^{N+n_{3}+1}\int_{0}^{\infty }%
\frac{y^{N+n_{3}}}{\left( y+1\right) ^{N+n_{3}+2}}dy,
\end{equation}%
and the desired bracket series of bracket is 
\begin{equation*}
\int_{0}^{X}t^{N+n_{3}}\;dt=\frac{X^{N+n_{3}+1}}{\Gamma \left(
N+n_{3}+2\right) }\sum\limits_{n_{4}}\sum\limits_{n_{5}}\phi _{n_{4}}\phi
_{n_{5}}\langle N+n_{3}+2+n_{4}+n_{5}\rangle \langle N+n_{3}+n_{4}+1\rangle.
\end{equation*}
\noindent
The final bracket series for $D_{N}(\alpha ,X)$ is 
\begin{equation}
\begin{array}{ll}
D_{N}\left( \alpha ,X\right) = & NX\sum\limits_{n_{1}}...\sum\limits_{n_{5}}%
\phi _{n_{1}} \cdots \phi _{n_{5}}\;\left( -1\right) ^{n_{2}-n_{3}}\frac{%
n_{1}^{n_{3}}}{\Gamma \left( N+n_{3}+2\right) }\alpha ^{n_{2}}X^{n_{3}} \\
&  \\
& \times \langle 1+n_{1}+n_{2}\rangle \langle N+n_{3}+2+n_{4}+n_{5}\rangle
\langle N+n_{3}+n_{4}+1\rangle.%
\end{array}
\label{eqq9}
\end{equation}

An expression for the integral \eqref{eqq3} is now obtained from 
\eqref{eqq9}. The method of brackets yields  \texttt{four} different series: 

\begin{eqnarray}
S_{1}& = & -\frac{NX}{\alpha }\sum\limits_{n_{1}\geq 0}\sum\limits_{n_{2}\geq 0}%
\frac{\Gamma \left( N+1+n_{2}\right) }{\Gamma \left( N+2+n_{2}\right) }\frac{%
n_{1}^{n_{2}}}{n_{2}!}\left( \frac{1}{\alpha }\right) ^{n_{1}} 
X^{n_{2}}, \label{series-1} \\
S_{2}& = & NX\sum\limits_{n_{1}\geq 0}\sum\limits_{n_{2}\geq 0}\left( -1\right)
^{n_{2}}\frac{\Gamma \left( N+1+n_{2}\right) }{\Gamma \left(
N+2+n_{2}\right) }\frac{\left( 1+n_{1}\right) ^{n_{2}}}{n_{2}!}\alpha^{n_{1}}X^{n_{2}}, \label{series-2} \\
S_{3}& = & \frac{N}{X^{N}}\sum\limits_{n_{1}\geq 0}\sum\limits_{n_{2}\geq 0}\left(
-1\right) ^{n_{2}}\frac{\Gamma \left( N+1+n_{2}\right) }{\Gamma \left(
1-n_{2}\right) }\frac{\left( 1+n_{1}\right) ^{-1-N-n_{2}}}{n_{2}!}
\frac{\alpha^{n_{1}} }{X^{n_{2}}}, \\
S_{4}& = & (-1) ^{N}\frac{N}{X^{N} \alpha }\sum\limits_{n_{1}\geq
0}\sum\limits_{n_{2}\geq 0}\frac{\Gamma \left( N+1+n_{2}\right) }{\Gamma
\left( 1-n_{2}\right) }\frac{n_{1}^{-N-1-n_{2}}}{n_{2}!}
\left(\frac{1}{\alpha}\right)^{n_{1}}  X^{n_{2}}.
\end{eqnarray}

The influence of the parameter $\alpha $ is discussed first, because in addition to
parameter $X$, it allows to discriminate different series. The four solutions $S_{j}$ are 
power series in $\alpha$ or $1/\alpha$, so that 
$S_{1}$ and $S_{4}$ are expansions in $\alpha \rightarrow \infty$ and $S_{2}$ and $S_{3}$ are expansions
in $\alpha \rightarrow 0$. The same situation occurs with respect to the parameter $X$. Each series represents the integral 
\eqref{eqq3}.  Their analysis is described next.

\begin{enumerate}
\item The series $S_{4}$  must be neglected because the term with $n_{1} = 0$ diverges.

\item The series $S_{3}$  is naturally truncated at  $n_{2}=0$. Since this index is associated to the  powers of  $X^{-1}$, it 
represents an asymptotic approximation for case $X >> 1$. A detailed study including
condition $\alpha\rightarrow 1$ yields:

\begin{equation}
S_{4}  \approx \frac{N\Gamma \left( N+1\right) }{X^{N}}\sum\limits_{n_{1}\geq 0}\frac{1}{\left( 1+n_{1}\right) ^{N+1}} 
 = \frac{N\Gamma \left( N+1\right) }{X^{N}}\zeta(N+1),
\end{equation}
where $\zeta(s)$ is the  Riemann zeta function.

\item The series $S_{1}$ and $S_{2}$ are both convergent as power series in $X$. Both are 
expressions for $D_{N}(\alpha ,X)$, but it turns out that they are equivalent.
 \end{enumerate}

\subsection{Analysis of the expressions obtained above}

\subsubsection{$S_{1}$ as solution} Rearranging the defining series produces a hypergeometric representation:

\begin{eqnarray}
S_{1} & = & - \frac{NX}{\alpha }\sum\limits_{n_{1}\geq 0}
\left( \alpha^{-1}\right) ^{n_{1}}\sum\limits_{n_{2}\geq 0}\frac{\Gamma \left(
N+1+n_{2}\right) }{\Gamma \left( N+2+n_{2}\right) }\frac{\left(
Xn_{1}\right) ^{n_{2}}}{n_{2}!} \\
& = & -\frac{NX}{\alpha \left( N+1\right) }\sum\limits_{n_{1}\geq 0}
\left(\alpha ^{-1}\right) ^{n_{1}}
\pFq11{N+1}{N+2}{Xn_{1}}.  \nonumber 
\end{eqnarray}
\noindent
The previous expression may be written as 
\begin{equation}
S_{1}=  -\left( \frac{N}{N+1}\right) \frac{X}{\alpha } -
\left( \frac{N}{N+1} \right)
\frac{X}{\alpha }\sum\limits_{n_{1}\geq 1}\left( \alpha ^{-1}\right)^{n_{1}} 
\pFq11{N+1}{N+2}{Xn_{1}},
\label{eqq1}
\end{equation}
where hypergeometric function $_{1}F_{1}$ is the Kummer function. Now use 
\begin{equation}
\pFq11{n}{n+1}{-Z} = \frac{n}{Z^{n}} \gamma(n,Z),
\label{1F1}
\end{equation}
where $\gamma( n,Z) $ is the incomplete Gamma function defined by the integral representation
\begin{equation}
\gamma(n,Z) =\int_{0}^{Z}t^{n-1}e^{-t}\;dt.
\end{equation}

In the important special case of  $n\in\mathbb{N}$, 
the function $\gamma(n,Z)$ can be written as a finite sum
\begin{equation}
\gamma \left( n,Z\right) =\Gamma \left( n\right) \left[ 1-e^{-Z}\sum
\limits_{k=0}^{n-1}\frac{Z^{k}}{k!}\right] ,
\end{equation}
and then
\begin{equation}
\pFq11{n}{n+1}{-Z}  =\frac{\Gamma \left( n+1\right) }{Z^{n}}\left[
1-e^{-Z}\sum\limits_{k=0}^{n-1}\frac{Z^{k}}{k!}\right] .  \label{eqq2}
\end{equation}

The formula \eqref{1F1}  is now  transformed to 
\begin{eqnarray*}
\pFq11{N+1}{N+2}{Xn_{1}} & = &  \frac{\Gamma \left( N+2\right) }{%
\left( -Xn_{1}\right) ^{N+1}}\left[ 1-e^{Xn_{1}}\sum\limits_{k=0}^{N}\frac{%
\left( -Xn_{1}\right) ^{k}}{k!}\right] \\
& = & (-1)^{N+1}\frac{\left( N+1\right) \Gamma \left( N+1\right) }
{X^{N+1}n_{1}^{N+1}}
\left[ 1-e^{Xn_{1}}\sum\limits_{k=0}^{N}\frac{\left(
-Xn_{1}\right) ^{k}}{k!}\right], \nonumber
\end{eqnarray*}
and the series $S_{1}$ can be  written as 
\begin{equation*}
S_{1}=  -\left( \frac{N}{N+1}\right) \frac{X}{\alpha } + \left( -1\right)
^{N}\frac{N\Gamma \left( N+1\right) }{X^{N}\alpha }\sum\limits_{n_{1}\geq 1}%
\frac{\left( \alpha ^{-1}\right) ^{n_{1}}}{n_{1}^{N+1}}\left[
1-e^{Xn_{1}}\sum\limits_{k=0}^{N}\frac{\left( -Xn_{1}\right) ^{k}}{k!}\right].
\end{equation*}
\noindent
After some algebraic manipulations, the previous expression is written as 

\begin{eqnarray*}
S_{1} & =& -\left( \frac{N}{N+1}\right) \frac{X}{\alpha }+\left( -1\right) ^{N}%
\frac{N\Gamma \left( N+1\right) }{X^{N}\alpha } \times \left[
\sum\limits_{n_{1}\geq 1}\frac{\left( \alpha ^{-1}\right) ^{n_{1}}}{%
n_{1}^{N+1}}-\sum\limits_{n_{1}\geq 1}\frac{\left[ \frac{e^{X}}{\alpha }%
\right] ^{n_{1}}}{n_{1}^{N+1}}\sum\limits_{k=0}^{N}\frac{\left(
-Xn_{1}\right) ^{k}}{k!}\right] \\
& &  \\
& =& -\left( \frac{N}{N+1}\right) \frac{X}{\alpha }+\left( -1\right) ^{N}\frac{%
N\Gamma \left( N+1\right) }{X^{N}\alpha } \times \left[ \sum\limits_{n_{1}%
\geq 1}\frac{\left( \alpha ^{-1}\right) ^{n_{1}}}{n_{1}^{N+1}}%
-\sum\limits_{k=0}^{N}\frac{\left( -X\right) ^{k}}{k!}\sum\limits_{n_{1}\geq
1}\frac{\left[ \frac{e^{X}}{\alpha }\right] ^{n_{1}}}{n_{1}^{N+1-k}}\right].%
\end{eqnarray*}

The polylogarithm function \cite{gradshteyn-2015a}, defined by the series 
\begin{equation}
\texttt{Li}_{s}\left( x\right) =\sum\limits_{k\geq 1}\frac{x^{k}}{k^{s}},
\end{equation}
\noindent
is now used to obtain an expression for the Debye function $D_{N}(\alpha,X)$ in the 
form 
\begin{eqnarray}
& & \label{eqq4} \\
D_{N}\left( \alpha ,X\right) & = & -\left( \frac{N}{N+1}\right) \frac{X}{\alpha} \nonumber   \\
& + & \left( -1\right) ^{N}\frac{N\Gamma \left( N+1\right) }{X^{N}\alpha }
\times \left[ \texttt{Li}_{N+1}\left( \alpha ^{-1}\right)
-\sum\limits_{k=0}^{N}\texttt{Li}_{N+1-k}\left( \frac{e^{X}}{\alpha }\right)
\frac{\left( -X\right) ^{k}}{k!}\right]. \nonumber 
\end{eqnarray}

This formula was first presented in \cite{dubinov-2008a}. In addition to this
representation, the method of brackets produces a new expression for the Debye function  using $S_{2}$.

\subsubsection{The series $S_{2}$. A new solution} As in the computation of $S_{1}$, the series defining $S_{2}$ can be 
written as a sum of values of the  incomplete Gamma function:

\begin{eqnarray}
S_{2} & = & NX\sum\limits_{n_{1}\geq 0}\sum\limits_{n_{2}\geq 0}\alpha ^{n_{1}}%
\frac{\Gamma \left( N+1+n_{2}\right) }{\Gamma \left( N+2+n_{2}\right) }\frac{%
\left( -X\right) ^{n_{2}}\left( 1+n_{1}\right) ^{n_{2}}}{n_{2}!} \\
& = & \frac{N}{N+1}X\sum\limits_{n_{1}\geq 0}\alpha ^{n_{1}}\;_{1}F_{1}\left(
\left.
\begin{array}{c}
N+1 \\
N+2%
\end{array}%
\right\vert -\left( 1+n_{1}\right)X \right)  \nonumber \\
& = & \frac{N}{X^{N}}\sum\limits_{n_{1}\geq 0}\frac{\alpha ^{n_{1}}}{\left(
1+n_{1}\right) ^{N+1}}\;\gamma \left( N+1,\left( 1+n_{1}\right)X \right), \nonumber 
\end{eqnarray}%
and using  \eqref{eqq2}, this becomes 
\begin{equation}
S_{2}=  \frac{N\Gamma \left( N+1\right) }{X^{N}\alpha } \times \left[
\sum\limits_{n_{1}\geq 0}\frac{\alpha ^{n_{1}+1}}{\left( 1+n_{1}\right)
^{N+1}}-\sum\limits_{k=0}^{N}\frac{X^{k}}{k!}\sum\limits_{n_{1}\geq 0}\frac{%
\left[ \alpha e^{-X}\right] ^{n_{1}}}{\left( 1+n_{1}\right) ^{N+1-k}}\right].
\end{equation}%
Proceeding as in the previous case, the Debye function $D_{N}(\alpha ,X)$ is now
\begin{equation}
D_{N}( \alpha ,X) =\frac{N\Gamma \left( N+1\right) }{X^{N}\alpha }%
\left[ \texttt{Li}_{N+1}\left( \alpha \right) -\sum\limits_{k=0}^{N}\texttt{Li}_{N+1-k}\left( \alpha e^{-X}\right) \frac{X^{k}}{k!}\right].  \label{eqq5}
\end{equation}

In summary, the method of brackets has produced two equivalent formulations of the representation of the 
Debye function given in \cite{dubinov-2008a}. The first one in \eqref{eqq3}, reproducing the solution 
presented in \cite{dubinov-2008a} and  a second
expression given in  \eqref{eqq5}. This is a new representation for $D_{N}(\alpha,X)$.

\section{Application : Debye Model and heat capacity in solids}
\label{sec-4}

An important topic in solid state physics is the 
determination of heat capacity using quantum treatments \cite{debye-1912a,einstein-1907a}. The integral expression
 (\ref{DebyeFunction}) is associated to this problem through a model proposed
by Debye \cite{debye-1912a}. According to this model, the internal energy in solids
is given as a function of the absolute temperature $T$, by
\begin{equation}
U=3Nk_{B}T D_{3}\left( \frac{\Theta _{D}}{T}\right),  \label{eqq8}
\end{equation}
with the usual notation  for Debye functions, i.e $%
D_{3}\left(\frac{\Theta _{D}}{T}\right) =D_{3}\left(1,\frac{\Theta _{D}}{T}\right)$. Here $k_{B}$ is the Boltzmann constant, 
$\Theta _{D}$ is called the \textit{Debye temperature} and $N$ the number of particles in the system.

Using \eqref{eqq4}, and with the notation $u = \Theta_{D}/T$, the Debye function is 
\begin{multline}
 \label{eqq6} 
D_{3}(u)  =  -\frac{3u}{4} -\frac{24}{u^{3}}
\zeta(4) \\
+\frac{24}{u^{3}}\left[
\texttt{Li}_{4}\left(e^{u} \right) -u \texttt{Li}_{3}\left( e^{u} \right)
     +\tfrac{1}{2}u^{2}\texttt{Li}_{2}\left( e^{u} \right) -\tfrac{1}{6}u^{3}\texttt{Li}_{1}\left( e^{u}\right) 
\right], 
\end{multline}
where $\zeta(4)=\pi^{4}/90$. The expressions \eqref{eqq6} and \eqref{eqq7} are analytical expressions for the Debye
functions $D_{3}\left( \frac{\Theta _{D}}{T}\right)$. These complement the original integral representation \eqref{debye-1}.

The analysis of \eqref{eqq6} as $T \rightarrow 0$ is not easy to obtain directly from here. On
the other hand, the new expression \eqref{eqq5} permits such an analysis.  To describe this procedure and with 
the same notation as before, start with 
\begin{multline}
\label{eqq7}
D_{3}(u) =  \frac{18}{u^{3}} \zeta(4) 
-\frac{18}{u^{3}}\texttt{Li}_{4}\left( e^{-u} \right)  \\ -
\frac{18}{u^{2}}\texttt{Li}_{3}\left( e^{-u} \right)  
 -\frac{9}{u} \texttt{Li}_{2}\left( e^{-u} \right) -3\texttt{Li}_{1}\left( e^{-u}\right) .
\end{multline}
\noindent
This is described next. 

\subsection{Asymptotic limits}

The classical approach to study limiting behavior of these functions is to conduct approximations for the 
integral representations, valid in some specific limits (high and low temperatures). These 
limits can now be studied directly from analytical expressions presented here. The new formulae 
presented here  permit the analysis of limiting  high and
low temperatures. This study reproduces the results of \cite{huang-1975a,mcquarrie-1975a}:

\bigskip

\begin{itemize}
\item At $T\rightarrow \infty $,
\end{itemize}

\begin{equation}
D_{3}\left( \frac{\Theta _{D}}{T}\right) \approx  1-\frac{3}{8}\frac{\Theta
_{D}}{T}+\frac{1}{20}\left( \frac{\Theta _{D}}{T}\right) ^{2} -\frac{1}{1680}%
\left( \frac{\Theta _{D}}{T}\right) ^{4}+O\left( T^{-6}\right) .%
\end{equation}

\begin{itemize}
\item At $T\rightarrow 0$
\end{itemize}
\begin{equation}
D_{3}\left( \frac{\Theta _{D}}{T}\right) \approx \frac{18}{\left( \frac{%
\Theta _{D}}{T}\right) ^{3}}\mathbf{\zeta }_{4}.
\end{equation}

In the analysis of this last formula, the behavior of the polylogaritmic function 
$\texttt{Li}_{n}\left(e^{-\Theta _{D}/T} \right) \ll 1$ as $T \rightarrow 0$ is used.  This can be seen from 
the power series expansion
\begin{equation}
\texttt{Li}_{n}\left( e^{-\frac{\Theta _{D}}{T}}\right) =  e^{-\Theta/T} + 
\frac{1}{2^{n}}e^{-2 \Theta _{D}/T} +\frac{1}{3^{n}}e^{-3\Theta _{D}/T}+ \ldots 
\end{equation}%
and this contribution is negligible in relation to $18 \zeta(4)\left( \frac{\Theta _{D}}{T}\right)^{-3}$. With
these approximations, the internal energy satisfies 

\begin{itemize}
\item For $T\rightarrow \infty $
\end{itemize}
\begin{equation}
U\approx 3Nk_{B}T-\frac{9}{8}Nk_{B}\Theta _{D}+\frac{3}{20}Nk_{B}\left(
\frac{\Theta _{D}^{2}}{T}\right) -\frac{1}{560}Nk_{B}\left( \frac{\Theta
_{D}^{4}}{T^{3}}\right) .
\end{equation}

\begin{itemize}
\item For $T\rightarrow 0$
\end{itemize}
\begin{equation}
U\approx \frac{3}{5}\frac{\pi ^{4}}{\Theta _{D}^{3}}Nk_{B}T^{4}.
\end{equation}%
These are in agreement with the result cited in the literature \cite{huang-1975a,mcquarrie-1975a}.

\subsection{Heat capacity}

This is computed using  $c_{V}=\left( \frac{\partial U}{\partial T}\right) _{V}$. The limiting behaviors are 

\begin{itemize}
\item For $T\rightarrow \infty $
\end{itemize}
\begin{equation}
c_{V}\approx 3Nk_{B}-\frac{3}{20}Nk_{B}\left( \frac{\Theta _{D}}{T}\right)
^{2}+\frac{3}{560}Nk_{B}\left( \frac{\Theta _{D}}{T}\right) ^{4}+O\left(
T^{-6}\right) .
\end{equation}

\begin{itemize}
\item For $T\rightarrow 0$
\end{itemize}
\begin{equation}
c_{V}\approx \frac{12\pi ^{4}}{5}\left( \frac{T}{\Theta _{D}}\right)
^{3}Nk_{B}.
\end{equation}

The analytical expressions for the Debye functions presented here produce results valid for arbitrary temperatures.  Using 
\eqref{eqq8}, and with the notation $u = \Theta_{D}/T$,  the value $c_{V}$ is given by 
\begin{eqnarray*}
\quad c_{V} & = & -\frac{12}{5}\pi ^{4}Nk_{B}u^{-3}+216Nk_{B}u^{-3}\texttt{Li}_{4}\left(
e^{u} \right)  -216Nk_{B}u^{-2}\texttt{Li}_{3}\left( e^{u} \right) \\
& & \quad +108Nk_{B} u^{-1}
\texttt{Li}_{2}\left( e^{u} \right)   -36Nk_{B}\texttt{Li}_{1}\left( e^{u} \right)
+9Nk_{B}u  \left( \frac{e^{u}}{{1-e^{u} }} \right), \nonumber 
\end{eqnarray*}
or using the new solution given in \eqref{eqq7}, 
\begin{eqnarray*}
c_{V} & = & \frac{12}{5}\pi ^{4}Nk_{B}u^{-3} -216Nk_{B}u^{-3}\texttt{Li}_{4}\left(
e^{-u}  \right)   -216Nk_{B}u^{-2}\texttt{Li}_{3}\left(e^{-u}\right)  \nonumber \\
& & -108Nk_{B}u^{-1} 
\texttt{Li}_{2}\left( e^{-u} \right) 
 -36Nk_{B}\texttt{Li}_{1}\left( e^{-u} \right)
-9Nk_{B}u  \left( \frac{e^{-u}}{1-e^{-u}} \right). \nonumber 
\end{eqnarray*}

\section{Conclusions}
\label{sec-5}

Analytic expressions for the Debye functions have been produced using the method of brackets. These 
expression differ from the classical integral representations and they involve sums of the polylogarithm function. One of the
 results presented here reproduces formulas developed in \cite{dubinov-2008a}.

The new expressions obtained here provide an efficient way to evaluate high and low 
temperature behavior.  

\medskip 

\noindent
\textbf{Acknowledgments}. 
The work of A.V. was supported by FONDECYT (Chile) under Grant No. 1141280, 
CONICYT (Chile) Research Project No. 7912010025 and FONDECYT (Chile) under grant No. 1180753.
I.K. was supported in part by Fondecyt (Chile) Grants Nos. 1040368, 1050512 and 1121030, by DIUBB (Chile) Grant Nos. 102609,  
GI 153209/C  and GI 152606/VC.

\end{document}